\RequirePackage{ifpdf}
\ifpdf 
\documentclass[pdftex]{sigma}
\else
\documentclass{sigma}
\fi

\begin{document}

\allowdisplaybreaks

\renewcommand{\PaperNumber}{043}

\FirstPageHeading

\renewcommand{\thefootnote}{$\star$}

\ShortArticleName{Riccati and Ermakov Equations in Quantum Theory}

\ArticleName{Riccati and Ermakov Equations in Time-Dependent\\ and
  Time-Independent Quantum Systems\footnote{This paper is a
contribution to the Proceedings of the Seventh International
Conference ``Symmetry in Nonlinear Mathematical Physics'' (June
24--30, 2007, Kyiv, Ukraine). The full collection is available at
\href{http://www.emis.de/journals/SIGMA/symmetry2007.html}{http://www.emis.de/journals/SIGMA/symmetry2007.html}}}

\Author{Dieter SCHUCH}

\AuthorNameForHeading{D.~Schuch}

\Address{Institut f\"ur Theoretische Physik, J.W. Goethe--Universit\"at,\\ Max-von-Laue-Str.~1, D-60438 Frankfurt am Main, Germany}
\Email{\href{mailto:schuch@em.uni-frankfurt.de}{schuch@em.uni-frankfurt.de}}

\ArticleDates{Received December 28, 2007, in f\/inal form May 07,
2008; Published online May 12, 2008}

\Abstract{The time-evolution of the maximum and the width of exact analytic
wave packet (WP) solutions of the time-dependent Schr\"odinger equation (SE)
represents the particle and wave aspects, respectively, of the quantum system. The dynamics
of the maximum, located at the mean value of position, is governed by
the Newtonian equation of the corresponding classical problem. The width, which
is directly proportional to the position uncertainty, obeys a complex
nonlinear Riccati equation which can be transformed into a real nonlinear
Ermakov equation. The coupled pair of these equations yields a
dynamical invariant which plays a key role in our investigation. It can be
expressed in terms of a  complex variable that linearizes the Riccati
equation. This variable also provides the time-dependent parameters that
characterize the Green's function, or Feynman kernel, of the corresponding
problem. From there, also the relation between the classical and quantum
dynamics of the systems can be obtained. Furthermore, the close connection
between the Ermakov invariant and the Wigner function will be
shown. Factorization of the dynamical
invariant allows for compari\-son with creation/annihilation operators and
supersymmetry where the partner potentials fulf\/il (real) Riccati
equations. This provides the link to a nonlinear formulation of
time-independent quantum mechanics in terms of an Ermakov equation for the
amplitude of the stationary state wave functions combined with a conservation
law. Comparison with SUSY and the time-dependent problems concludes our analysis.}


\Classification{37J15; 81Q05; 81Q60; 81S30}

\section{Introduction}

Participating in a conference on nonlinear mathematical physics taking place
in Kiev, one can f\/ind more colleagues than usual who are familiar with the
name Ermakov; some are even able to tell where one can f\/ind traces of him from
the time he used to work at the university in Kiev. This period was, however,
before quantum mechanics was established and Vasili Petrovich Ermakov
(1845--1922) died already four years before Schr\"odinger published his
celebrated equation \cite{1}. So, what can be his relation with quantum theory?
The second name in the title of this paper also seems to be surprising in this
context because Jacopo Riccati (1676--1754) was born actually 250 years prior
to Schr\"odinger's publication! The surprising answer to our question is that
Ermakov's and Riccati's contributions are not only closely connected with
quantum theory in the form established by Schr\"odinger and Heisenberg, but
also with later formulations due to Feynman and Wigner. Therefore, their work
plays a central role in f\/inding underlying connections amongst all these
dif\/ferent formulations.

One way of showing how Riccati and Ermakov enter the formalism of quantum theory
is the study of cases where exact analytic Gaussian wave packet (WP)
solutions of the time-dependent Schr\"odinger equation (SE) exist, in
particular, the harmonic oscillator (HO)
and the free motion. Therefore, in the following, we will restrict the
discussion of the time-dependent problems to the (one-dimensional) HO (where the frequency $\omega$ may also be
time-dependent) and the free motion is obtained in the limit $\omega
\rightarrow 0$. In Section~2, it will be shown how a (complex) nonlinear Riccati equation that
governs the dynamics of a typical quantum ef\/fect, namely, the position
uncertainty, emerges from the WP solution of the time-dependent SE. This
Riccati equation can be transformed into a (real) nonlinear Ermakov equation
that, together with the classical Newtonian equation of the system, provides a
corresponding dynamical invariant, the so-called Ermakov invariant, that will
play an important role in our discussion. As shown in Section~3, the linearization
of the Riccati equation can be accomplished by introducing new complex
variables. These variables are connected with the time-dependent parameters
that characterize the time-dependent Green's function, also called Feynman
kernel, of the corresponding problem. They also provide the key to
understanding the relations between the dynamics of the classical and quantum
aspects of the system; in particular, the importance of the initial position
uncertainty will become obvious. Furthermore, in Section~4, the close
relationship between the Ermakov invariant and the Wigner function will be
shown. The link to the occurrence of Riccati and Ermakov equations in
time-independent quantum mechanics is obtained in Section~5. Factorization of
the Ermakov invariant allows for a comparison with creation and annihilation
operators as well as their generalizations used in supersymmetry (SUSY). In
the latter case, the supersymmetric partner potentials fulf\/il $real$ Riccati
equations. This shows similarities with a~nonlinear formulation of
time-independent quantum mechanics by Reinisch~\cite{2}, where he obtains an
Ermakov equation (with spatial derivatives instead of temporal ones) for the
amplitude of the stationary state wave functions. This Ermakov equation is
related with a kind of {\it complexification} of the Riccati equation
appearing in SUSY. Finally, in Section~6, the main results will be summarized
and possible generalizations mentioned.

\section{Dynamics of Gaussian wave packets and Ermakov invariant}

Starting point of our investigation is the time-dependent SE for the HO with frequency $\omega (t)$,
\begin{gather}
 i\hbar \frac{\partial}{\partial t} \Psi_{\rm WP} = \left\{ -\frac{\hbar^{2}}{2m} \frac{\partial^{2}}{\partial x^{2}} + \frac{m}{2} \omega^2(t) x^2 \right\} \Psi_{\rm WP}
\label{ric01}
\end{gather}
that possesses Gaussian WP solutions of the type
\begin{gather}
\Psi_{\rm WP}(x,t) = N(t) \exp \left\{ i\left[y(t) \tilde{x}^{2}+\frac{\langle p\rangle }{\hbar}\tilde{x}+K(t)\right]
\right\} ,
\label{ric02}
\end{gather}
where $y(t)=y_{\rm R}(t)+i y_{\rm I}(t)$, $\tilde{x}=x-\langle x\rangle=x-\eta (t)$ (i.e., $\langle x\rangle = \eta (t)$ is the classical trajectory calculated as mean value $\int \Psi^\ast_{\rm WP} x \Psi_{\rm WP} dx = \langle x\rangle$), $\langle p\rangle =m\dot{\eta}$ is the classical momentum and $N(t)$ and $K(t)$ are purely time-dependent terms that are not relevant for the following discussion.

Inserting the WP \eqref{ric02} into equation~\eqref{ric01} yields the equations of motion for $\eta
(t)$ and $y(t)$. The equation for the WP maximum, located at $x=\eta (t)$, is just the classical equation of motion
\begin{gather}
\ddot{\eta} + \omega^2(t){\eta}   = 0,
\label{ric03}
\end{gather}
where overdots denote time derivatives.

The equation of motion for the complex quantity $y(t)$ that is connected with
the WP width and, thus, the position uncertainty, is given by the complex Riccati-type equation
\begin{gather}
 \frac{2\hbar}{m}\dot{y}+ \left(\frac{2\hbar}{m}y\right)^{2}+\omega^{2}(t) = 0 .	
\label{ric04}
\end{gather}
Equation \eqref{ric04} can be separated into real and imaginary parts,
\begin{gather}
 {\rm Im}: \ \ \frac{2\hbar}{m}\dot{y}_{\rm I}+ 2\left(\frac{2\hbar}{m}y_{\rm I}\right) \left(\frac{2\hbar}{m}y_{\rm R}\right)  = 0 ,	
\label{ric05}
\\
 {\rm Re}: \ \ \frac{2\hbar}{m}\dot{y}_{\rm R} + \left(\frac{2\hbar}{m}y_{\rm R}\right)^2 - \left(\frac{2\hbar}{m}y_{\rm I}\right)^2 + \omega^{2}(t) = 0.	
\label{ric06}
\end{gather}
The real part $y_{\rm R} (t)$ can be eliminated from equation~\eqref{ric06} by solving equation~\eqref{ric05} for
$y_{\rm R}$ and inserting the result into equation~\eqref{ric06}.

It is useful to introduce a new variable $\alpha (t)$ that is connected with $y_{\rm I} (t)$ via
\begin{gather}
\frac{2\hbar}{m}y_{\rm I} = \frac{1}{\alpha^{2}(t)} ,	
\label{ric07}
\end{gather}
where $\alpha (t)$ is directly proportional to the WP width, or position
uncertainty, i.e., $\alpha = \!\left(\frac{m}{2\hbar y_{\rm I}}\right)^{1/2}\! = \left(\frac{2m
\langle \tilde{x}^2(t)\rangle }{\hbar}\right)^{1/2}$ with $\langle \tilde{x}^2\rangle = \langle x^2\rangle -
\langle x \rangle^2$. Inserting this def\/inition \eqref{ric07} into equation~\eqref{ric05} shows that the real part of
$y(t)$ just describes the relative change in time of the WP width,
\begin{gather}
\frac{2\hbar}{m}y_{\rm R} = \frac{\dot{\alpha}}{\alpha} = \frac{1}{2}\frac{\frac{d}{dt}\langle \tilde{x}^2\rangle }{\langle \tilde{x}^2\rangle}.	
\label{ric08}
\end{gather}
Together with def\/inition \eqref{ric07}, this f\/inally turns equation~\eqref{ric06} into
\begin{gather}
\ddot\alpha + \omega^2(t)\alpha   = \frac{1}{\alpha^3},
\label{ric09}
\end{gather}
the so-called Ermakov equation.

It has been shown by Ermakov in 1880~\cite{3} that the system of
dif\/ferential equations \eqref{ric03} and~\eqref{ric09}, coupled via
the possibly time-dependent frequency $\omega$, leads to a dynamical
invariant that has been rediscovered by several authors in the
20th century~\cite{4},
\begin{gather}
I_{\rm L} = \frac{1}{2} \left[ \left( \dot{\eta} \alpha - \eta \dot{\alpha} \right)^2 + \left( \frac{\eta}{\alpha} \right)^2 \right]  =  \textrm{const}.
\label{ric10}
\end{gather}
It is straightforward to show that $\frac{d}{dt} I_{\rm L} = 0$; a proof following
Ermakov's method can be found in \cite{5}. The Ermakov invariant not only
depends on the classical variables $\eta (t)$ and $\dot{\eta} (t)$, but also
on the quantum uncertainty connected with $\alpha (t)$ and $\dot{\alpha} (t)$. Additional interesting insight into the relation between the variables $\eta$ and $\alpha$ can be obtained from a dif\/ferent treatment of the Riccati equation~\eqref{ric04}.

\section{Linearization of the complex Riccati equation,\\ Feynman kernel and
  quantum-classical connection}

Introducing a new {\it complex} variable $\lambda (t)$, the complex variable in
the Riccati equation \eqref{ric04} can be replaced by the logarithmic time-derivative of $\lambda$, i.e.,
\begin{gather}
 \left(\frac{2\hbar}{m}y\right) =  \frac{\dot{\lambda}}{\lambda},
\label{ric11}
\end{gather}
thus turning the nonlinear Riccati equation into the complex linear equation{\samepage
\begin{gather*}
\ddot{\lambda} + \omega^2(t){\lambda}   = 0,
\end{gather*}
looking exactly like equation~\eqref{ric03} for $\eta (t)$, which is not just by accident, as will be shown later.}

The {\it complex} variable $\lambda(t)$ can be written in polar coordinates as
well as in cartesian coordinates, i.e.,
\begin{gather*}
\lambda = \alpha e^{i\varphi} = u + i z.
\end{gather*}
The choice of the symbol $\alpha$ for the absolute value of $\lambda$ is also not
coincidental, as will now be shown.

\subsection{Polar form of the complex linearization variable}

Writing relation \eqref{ric11} in polar coordinates yields
\begin{gather}
\frac{\dot{\lambda}}{\lambda} =  \frac{\dot\alpha}{\alpha} + i
            \dot \varphi  =  \left(\frac{2\hbar}{m}y\right).
\label{ric14}
\end{gather}
The real part already looks identical to the one given in equation~\eqref{ric08}. So, the absolute value of $\lambda$ would be (up to a constant factor) identical with the square root of the position uncertainty~$\langle \tilde{x}^2\rangle$, if
\begin{gather}
\dot\varphi = \frac{1}{\alpha^2}
\label{ric15}
\end{gather}
is fulf\/illed. This can easily be verif\/ied by inserting $\left(\frac{2\hbar}{m}y_{\rm R}
\right)$ and $\left(\frac{2\hbar}{m}y_{\rm I} \right)$, as given
in equation~\eqref{ric14}, into the imaginary part of the Riccati equation, thus turning
equation~\eqref{ric05} into
\begin{gather*}
\ddot\varphi + 2 \frac{\dot{\alpha}}{\alpha} \dot\varphi = 0 ,
\end{gather*}
in agreement with equation~\eqref{ric15}.
From equation~\eqref{ric06} for the real part then, again, the Ermakov equation is obtained as
\begin{gather*}
\ddot\alpha + \omega^2(t)\alpha = \alpha \dot\varphi^2  = \frac{1}{\alpha^3} .
\end{gather*}

\subsection{Cartesian form and Feynman kernel}

After the physical meaning of the absolute value of $\lambda$ in polar
coordinates and its relation to the phase angle via \eqref{ric15} have been clarif\/ied,
the interpretation of the cartesian coordinates $u$ and $z$ needs to be
ascertained. For this purpose, it can be utilized that the WP solution \eqref{ric02} at time $t$
can also be obtained with the help of an initial WP at, e.g., $t' = 0$ and a
time-dependent Green's function, also called time-propagator or Feynman
kernel, via
\begin{gather}
 \Psi_{\rm WP}(x,t) = \int dx' G(x,x',t,t'=0) \Psi_{\rm WP}\left(x',0\right) .
\label{ric18}
\end{gather}
For the considered Gaussian WP with initial distribution
\begin{gather}
\Psi_{\rm WP}\left(x',0\right) = \left(\frac{m\beta_0}{\pi \hbar}\right)^{1/4} \exp \left\{\frac{im}{2\hbar}\left[i \beta_0 x'^2 + 2 \frac{p_0}{m} x' \right]\right\} ,
\label{ric19}
\end{gather}
where $\beta_0=\frac{\hbar}{2m \langle \tilde{x}^2\rangle _0}=\frac{1}{\alpha_0^2}$ and $p_0=\langle p\rangle (t=0)$, the Green's function can be written as
\begin{gather}
 G \left(x,x',t,0\right) = \left(\frac{m}{2\pi i \hbar \alpha_0 z}\right)^{1/2} \exp \left\{\frac{im}{2\hbar}\left[ \frac{\dot{z}}{z} x^2 - 2 \frac{x}{z}\left(\frac{x'}{\alpha_0}\right) + \frac{u}{z}\left(\frac{x'}{\alpha_0}\right)^2 \right]\right\} .
\label{ric20}
\end{gather}
Since in the def\/inition of $\Psi_{\rm WP}(x,t)$, according to equation~\eqref{ric18}, only $G$
actually depends on~$x$ and~$t$, the kernel~$G$, as def\/ined in equation~\eqref{ric20}, must
also fulf\/il the time-dependent SE. Inser\-ting~\eqref{ric20} into the SE~\eqref{ric01} shows
that $z(t)$ and $u(t)$ not only fulf\/il the same equation of motion as $\eta
(t)$ and $\lambda(t)$ but, in addition, are also uniquely connected via the relation
\begin{gather}
\dot{z} u - \dot{u} z = 1 .
\label{ric21}
\end{gather}
Expressing $u$ and $z$ in polar coordinates according to $u = \alpha \cos
\varphi$ and $z = \alpha \sin \varphi$ shows that this coupling is identical
to relation \eqref{ric15} that connects the amplitude and phase of $\lambda$. From
equation~\eqref{ric21} it also follows that, e.g., with the knowledge of $z$, the quantity $u$ can
be calculated~as
\begin{gather*}
u = -z \int _{}^{t} \frac{1}{z^2} dt' .
\end{gather*}
The last necessary step is to explicitly perform the integration in \eqref{ric18}, using \eqref{ric19} and \eqref{ric20}, to yield the WP solution in the form
\begin{gather*}
\Psi_{\rm WP} (x,t) =
         \left(\frac{m}{\pi\hbar}\right)^{1/4}
         \left(\frac{1}{ u + i z}\right)^{1/2} \exp
         \left\{\frac{im}{2\hbar}
         \left[ \frac{\dot{ z}}{ z} x^2 -
         \frac{(x-\frac{p_0\alpha_0}   {m}  z)^2}
        { z ( u + i  z)}\right] \right\}.
\end{gather*}

Comparison with the same WP, written in the form given in equation~\eqref{ric02}, shows that the relations
\begin{gather}
 z = \frac{m}{\alpha_0 p_0} \eta (t)
\label{ric24}
\end{gather}
and
\begin{gather*}
\frac{2\hbar}{m} y = \frac{\dot{ z}}{ z} -
     \frac{1}{ z\lambda} = \frac{\dot\lambda}{\lambda}
\end{gather*}
are valid, where $\lambda =  u + i z$ is identical to our linearization
variable and equation~\eqref{ric21} has been applied. So, from equation~\eqref{ric24}, it then follows that
the imaginary part of~$\lambda$ is, apart from a~constant factor, just the
particle trajectory.

The equivalence between deriving the time-dependent Green's function via a
Gaussian ansatz or via Feynman's path integral method has been mentioned in
\cite{6}, where also the relation to the Ermakov invariant is considered.

In conclusion, one can say that the complex quantity $\lambda$ contains the
particle as well as the wave aspects of the system. In polar coordinates, the absolute value $\alpha$
of $\lambda$ is directly connected with the quantum mechanical position
uncertainty; in cartesian coordinates, the imaginary part of $\lambda$ is
directly proportional to the classical particle trajectory $\eta$. Absolute
value and phase, or real and imaginary parts, of $\lambda$ are not independent of each other but uniquely connected via the conservation laws \eqref{ric15} and \eqref{ric21}, respectively.

\subsection{Quantum-classical connection}

Further insight into the relation between the classical and quantum
dynamics of the system can be gained by rewriting the invariant \eqref{ric10}, with the
help of equation~\eqref{ric24}, in terms of $z$ and $\dot{z}$ instead of $\eta$ and $\dot{\eta}$,
\begin{gather*}
 I_{\rm L}  = \frac12 \left(\frac{\alpha_0p_0}{m}\right)^2 \left[ \left( \dot{z}
 \alpha - z \dot \alpha\right)^2 + \left( \frac{z}{\alpha}\right)^2 \right]=
        \textrm{const} .
\end{gather*}
Since  $\frac{z}{\alpha} = \sin \varphi$, for $I_{\rm L}$ to be constant it is
necessary that $(\dot {z} \alpha - z \dot \alpha)^2 = \cos^2 \varphi  =
\left(\frac{u}{\alpha}\right)^2$ is valid. So, up to a $\pm$ sign, one obtains
\begin{gather}
u = \dot{z} \alpha^2 - z \dot \alpha
            \alpha = \alpha^2 \left(\dot{z} -
            \frac{\dot\alpha}{\alpha} z\right) .
\label{ric27}
\end{gather}
With this form of $u$, a certain {\it asymmetry} in the exponent of the Feynman
kernel \eqref{ric20}, namely $\frac{\dot{z}}{z}$ as coef\/f\/icient of $x^2$ compared to
$\frac{1}{\alpha^2_0} \frac{u}{z}$ as coef\/f\/icient of the initial quantity
$x'^2$, can be explained. According to \eqref{ric27}, $\frac{\dot{z}}{z}$ can be written as
\begin{gather}
 \frac{\dot{z}}{z} =
                    \frac{1}{\alpha^2(t)} \frac{u}{z} +
                    \frac{\dot\alpha}{\alpha} .
\label{ric28}
\end{gather}
This shows that, in the case when $\alpha$ is time-dependent, not
only $\alpha^2_0$ must be replaced by $\alpha^2(t)$, but also an
additional term $\frac{\dot\alpha}{\alpha}$ must be taken into
account. For constant $\alpha = \alpha_0$ or at $t = 0$, relation \eqref{ric28} reduces to
\begin{gather}
\frac{\dot{z}}{z} = \frac{1}{\alpha^2_0} \frac{u}{z} \qquad \textrm{or}
     \qquad u = \alpha^2_0 \dot{z} = \frac{m \alpha_0}{p_0} \dot{\eta} ,
\label{ric29}
\end{gather}
i.e., $u$ is simply proportional to $\dot{\eta}$. Note the explicit
occurrence of $\alpha_0$, the initial position uncertainty, because it has
important consequences for the time-dependence of $\alpha
(t)$. Inserting~$u$, as given in \eqref{ric29}, into $\alpha^2(t) = u^2 + z^2$ yields (with $v_0 = \frac{p_0}{m}$ and
$\beta_0 = \frac{1}{\alpha^2_0}$)
\begin{gather}
   \alpha^2(t) = \frac{\alpha^2_0}{v^2_0} \left[ \dot\eta^2
            + \beta^2_0 \eta^2 \right] = \frac{2m}{\hbar} \langle \tilde
            x^2\rangle .
\label{ric30}
\end{gather}
This shows that the {\it quantum mechanical uncertainty} of
position (at any time $t$) can be {\it expressed solely} in terms
of the {\it classical trajectory} $\eta (t)$ and the
{\it corresponding velocity} $\dot \eta (t)$, if the {\it initial
velocity $v_0$} {\it and} the {\it initial position uncertainty},
expressed by $\alpha_0$ are known.

This explains why Feynman's procedure \cite{7} of deriving his kernel based only on the $classical$
Lagrangian provides the correct time-evolution of the system since
the time-dependence enters only via the classical variables $\eta (t)$ and
$\dot{\eta} (t)$. However, the importance of the initial position uncertainty
$\alpha_0$ for the quantum dynamics should not be underestimated.

 The inf\/luence of the initial
uncertainty becomes clear if one inserts the expressions for $\eta
(t)$ and $\dot \eta (t)$ into equation~\eqref{ric30}.

a) For the free motion, one obtains with $\eta = v_0 t$, $\dot \eta =
v_0$:
\begin{gather*}
\alpha^2_{\rm fr}  = \alpha^2_0 \left[ 1+ (\beta_0 t)^2\right];
\end{gather*}

b) for the HO with $\eta = \frac{v_0}{\omega} \sin\omega t$ and
$\dot \eta = v_0 \cos \omega t$, equation~\eqref{ric30} yields
\begin{gather*}
\alpha^2_{\rm HO}  = \alpha^2_0 \left\{ \cos^2\omega t +
              \left(\frac{\beta_0}{\omega} \sin \omega t\right)^2
              \right\} .
\end{gather*}

Only if the initial state is the ground state, is $\beta_0 =
\frac{\hbar}{2m \langle \tilde x^2\rangle_0} = \frac{1}{\alpha^2_0} = \omega$ valid and, hence,
$\alpha =  \alpha_0$, i.e., the WP width remains constant; in all other cases it
oscillates. This oscillating WP width corresponds to the general solution of
the Riccati equation \eqref{ric04} and yields in the limit $\omega \rightarrow 0$ the
correct spreading WP width of the free motion WP,
\begin{gather*}
\lim_{\omega\to 0} \alpha^2_{\rm HO} (t) = \alpha^2_0 \left[1 +
        (\beta_0 t)^2\right] = \alpha^2_{\rm fr} (t) ,
\end{gather*}
whereas the WP usually presented, with constant width $\alpha_0 = \left(
  \frac{1}{\omega} \right)^{1/2}$, provides in this limit only a plane wave, no
WP!

The time-derivative of $\alpha^2$, or,
\begin{gather*}
 \dot\alpha \alpha = \frac{\alpha^2_0}{v^2_0} \dot\eta \left[
       \beta^2_0 \eta + \ddot \eta\right] = \frac{\alpha^2_0}{v^2_0} \dot\eta \left[
       \beta^2_0 \eta -  \frac{\partial}{\partial \eta} V (\eta) \right],
\end{gather*}
respectively, only vanishes if the term in square brackets is equal to
zero, which depends on~$ \frac{\partial}{\partial \eta} V (\eta)$. Therefore,
for $V = 0$ ($\ddot{\eta} = 0$), $\dot{\alpha} \ne 0$ is always valid; for the
HO ($\ddot{\eta} = - \omega^2 \eta$), $\dot{\alpha} = 0$ is only valid for
$\beta_0 =  \frac{1}{\alpha^2_0} = \omega$, otherwise, and in particular for
$\omega = \omega (t)$ (which describes,
e.g., the motion of an ion in a Paul trap \cite{8}), $\dot{\alpha} \ne
0$ always holds. So, obviously the initial value $\alpha_0$ of the position uncertainty
plays an important role in the qualitative behaviour of the quantum aspect of
the dynamics. A more detailed discussion of this problem can be found in~\cite{9}, also including dissipative ef\/fects.

\section{Ermakov invariant and Wigner function}

In the previous section, it has been shown how the real and imaginary parts
(in cartesian coordinates) of the complex variable $\lambda (t)$, that allows the
linearization of the complex Riccati equation~\eqref{ric04}, enter the Feynman kernel that describes the
transformation of an initial quantum state into a state at a later time $t$ as
time-dependent parameters. But, also in polar coordinates~$\alpha$ and~$\varphi$ (which are not independent of
each other but related via $\dot{\varphi} = \frac{1}{\alpha^2}$), $\lambda (t)$ is related, via the Ermakov invariant
\eqref{ric10}, with another description of quantum systems that shows close similarity
with the classical phase space description of dynamical systems, namely the
Wigner function. In order to show this connection the invariant~$I_{\rm L}$
shall be written explicitly in the form
\begin{gather*}
I_{\rm L} = \frac{1}{2} \left[ \left( \dot \alpha^2 + \frac{1}{\alpha^2} \right)
  \eta^2 - 2 \alpha \dot{\alpha} \eta \dot{\eta} + \alpha^2 \dot\eta^2 \right]
\end{gather*}
with terms bilinear in $\eta$ and $\dot{\eta}$. How the coef\/f\/icients of these
terms are  related with the quantum uncertainties, and thus with $\alpha$,
$\varphi$ and $\lambda$, follows from
\begin{gather}
\lambda \lambda^* =  \alpha^2 =  \frac{m}{2 \hbar y_{\rm I}} =  \frac{2m}{\hbar}
\langle \tilde x^2\rangle_{\rm L} ,
\label{ric36}
\\
\dot\lambda \dot\lambda^* = \dot \alpha^2 + \alpha^2
\dot\varphi^2 = \frac{\hbar}{m} \frac{y^2_{\rm R}+y^2_{\rm I}}{y_{\rm I}} = \frac{\hbar m}{2}
\langle \tilde p^2\rangle_{\rm L},
\label{ric37}
\\
 \frac{\partial}{\partial t}(\lambda \lambda^*) = 2 \dot \alpha \alpha  = 2
 \left(\frac{y_{\rm R}}{y_{\rm I}}\right) =  \frac{2}{\hbar} \langle [\tilde x, \tilde p]_+\rangle_{\rm L}
 = \frac{2}{\hbar} \langle \tilde x \tilde  p + \tilde p \tilde x\rangle_{\rm L} .
\label{ric38}
\end{gather}

With the help of these relations, the invariant takes the form
\begin{gather}
I_{\rm L} = \frac{1}{2} \left[ \dot\lambda \dot\lambda^* \eta^2
  - \frac{\partial}{\partial t}(\lambda \lambda^*)  \eta \dot{\eta} + \lambda \lambda^* \dot\eta^2 \right]
\nonumber \\
\phantom{I_{\rm L}}{}   = \frac{1}{m \hbar}\left[ \langle \tilde{p}^{2}\rangle _{\rm L}  \eta^{2} -
\langle [\tilde{x},\tilde{p}]_+\rangle _ {\rm L}  \eta (m \dot{\eta}) + \langle \tilde{x}^{2}\rangle_{\rm L}  (m \dot{\eta})^{2} \right].
\label{ric39}
\end{gather}

The connection with the Wigner function becomes obvious if one performs the
Wigner transformation \cite{10} of our WP \eqref{ric02} according to
\begin{gather*}
 W(x,p,t) = \frac{1}{2\pi\hbar} \int^{+\infty}_{-\infty} dq\,
        e^{ipq/\hbar} \Psi^*_{\rm WP} \left(x+\frac{q}{2}, t\right) \Psi_{\rm WP}
        \left(x-\frac{q}{2}, t\right)
\end{gather*}
yielding the Wigner function in the form
\begin{gather*}
 W(x,p,t) = \frac{1}{\pi\hbar} \exp \left\{-2 \left(\frac{y^2_{\rm I}+y^2_{\rm R}}{y_{\rm I}}
        \right) \tilde x^2 - \frac{\tilde p^2}{2\hbar^2 y_{\rm I}} +
        \frac{2}{\hbar}
        \left(\frac{y_{\rm R}}{y_{\rm I}}\right) \tilde x \tilde p\right\}.
\end{gather*}
Using relations \eqref{ric36}--\eqref{ric38}, this can be expressed as
\begin{gather}
W(x,p,t) = \frac{1}{\pi\hbar}\exp \left\{
                    -\frac{2}{\hbar^2} \left[ \langle \tilde p^2\rangle_{\rm  L} \tilde x^2 -
                    \langle [\tilde x, \tilde p]_+\rangle _{\rm L} \tilde x \tilde p  +
                    \langle \tilde x^2\rangle_{\rm  L} \tilde p^2\right]\right\}.
\label{ric42}
\end{gather}
In particular, at the origin of the phase space, i.e., for $x = 0$ and $p = 0$,
where $\tilde x^2 \rightarrow \eta ^2$, $\tilde p^2 \rightarrow
(m \dot{\eta})^2$ and $\tilde x \tilde p \rightarrow m \eta \dot{\eta}$, one obtains
\begin{gather*}
W(0,0,t) = \frac{1}{\pi \hbar} \exp \left\{ - \frac{2m}{\hbar} I_{\rm L} \right\} = \textrm{const}
\end{gather*}
with $ I_{\rm L}$ as given in \eqref{ric39} which fulf\/ils $ \frac{\partial}{\partial t}
W = 0$ since $ I_{\rm L}$ is an invariant. For $x \ne 0$  and $p \ne 0$, $W(x,p,t)$ takes the form \eqref{ric42} and the equation of motion is given,
as expected, by
\begin{gather*}
 \frac{\partial}{\partial t} W(x,p,t) = - \frac{p}{m}
        \frac{\partial W}{\partial x} + \frac{\partial V}{\partial
        x}\frac{\partial W}{\partial p},
\end{gather*}
i.e., a continuity equation for an incompressible medium, also called
Liouville equation in phase space.

Further details, also concerning the dif\/ferent forms to express the Ermakov
invariant in the exponent of the Wigner function and the physical
interpretation of these forms is given in~\cite{11}. Other attempts to
construct connections between the Ermakov invariant and the Wigner function
can be found in~\cite{6}.

\section{Riccati and Ermakov equations in time-independent\\ quantum mechanics}

\subsection{Factorization of the invariant and creation/annihilation operators}

Formal similarities between the Ermakov invariant and the algebraic treatment
of the HO using creation and annihilation operators (or complex normal modes, in
the classical case) and its generalization in the formalism of
SUSY can be found if $I_{\rm L}$ is written in a form that allows for factorization
\begin{gather*}
 I_{\rm L} = \frac12  \alpha^2  \left[ \left( \dot \eta  -  \frac{\dot
            \alpha}{\alpha}  \eta  \right)^2 +
            \left(\frac{\eta}{\alpha^2}\right)^2\right] = \frac12 \alpha^2 A A^*
\end{gather*}
with
\begin{gather}
A = \left( \dot \eta  -  \frac{\dot
            \alpha}{\alpha}  \eta  \right) - i \frac{1}{\alpha^2}
          \eta = \dot{\eta} - \left(\frac{2 \hbar}{m} y \right) \eta
\label{ric46}
\end{gather}
and
\begin{gather}
 A^* = \left( \dot \eta  -  \frac{\dot
            \alpha}{\alpha}  \eta  \right) + i \frac{1}{\alpha^2}
          \eta = \dot{\eta} - \left(\frac{2 \hbar}{m} y^* \right) \eta .
\label{ric47}
\end{gather}

For the HO with time-independent frequency $\omega_0 $ and constant WP width
$\alpha_0$, the real part of $ \left(\frac{2 \hbar}{m} y \right)$ vanishes
($\dot{\alpha} = 0)$ and the imaginary part is simply $\frac{1}{\alpha^2_0} =
\dot{\varphi} = \omega_0$, so
equations~\eqref{ric46} and~\eqref{ric47} turn into
\begin{gather}
A_0 =  \dot \eta - i \omega_0 \eta , \qquad A^*_0 =  \dot \eta + i \omega_0 \eta .
\label{ric48}
\end{gather}
These expressions are, up to constant factors, identical with the complex
normal modes or, in the quantized form, with the creation and
annihilation operators of the HO. In this context, it should be mentioned that
the Ermakov invariant has also been quantized, where $\alpha$ and $\dot{\alpha}$ remain
c-numbers, whereas position $\eta$ and momentum $m\dot{\eta}$ are, following the rules of
canonical quantization, replaced by the corresponding operators, i.e., $\eta
\rightarrow q_{\rm op} = q$, $m\dot{\eta} \rightarrow  p_{\rm op} =
\frac{\hbar}{i}\frac{\partial}{\partial q}$ (for details see, e.g.,~\cite{12}). Expressions~\eqref{ric48} would then turn into the operators
\begin{gather*}
A_{0,{\rm op}} =  \frac{1}{m} p_{\rm op}  - i \omega_0 q , \qquad A^*_{0,{\rm op}} =  \frac{1}{m} p_{{\rm op}}  + i \omega_0 q  .
\end{gather*}

For comparison, the Hamiltonian operator of the HO can be written as
\begin{gather*}
H_{{\rm op},{\rm HO}} =  \frac{1}{2m} p_{\rm op}^2 +  \frac{m}{2 } \omega_0^2 q^2 =\hbar
\omega_0 \left(\hat{b}^+ \hat{b}^- + \frac{1}{2} \right)
\end{gather*}
with the creation/annihilation operators
\begin{gather}
 \hat{b}^{\pm} = \mp i \, \sqrt{\frac{m}{2 \hbar \omega_0}} \left(\frac{p_{\rm op}}{m} \pm i
 \omega_0 q\right) = \sqrt{\frac{m}{2 \hbar \omega_0}} \left(\omega_0 q \mp  i  \frac{p_{\rm op}}{m}\right) .
\label{ric51}
\end{gather}
Comparison shows that
\begin{gather*}
\alpha_0 A_{0,{\rm op}} = -i  \sqrt{\frac{2 \hbar}{m}}  \hat{b}^- ,\qquad \alpha_0
A^*_{0,{\rm op}} = +i \sqrt{\frac{2 \hbar}{m}} \hat{b}^+.
\end{gather*}

So, $A$ and $A^*$ (or $A_{\rm op}$ and $A^*_{\rm op}$) are generalizations where the constant
factor $\pm i \omega_0$, in front of $q$, is replaced by the complex time-dependent
functions $ \left(\frac{2 \hbar}{m} y \right)$ or $\left(\frac{2 \hbar}{m} y^*
\right)$, respectively. The factorization of the dynamical invariant and the
relation to creation/annihilation operators is also discussed in~\cite{6}.

\subsection{Riccati and SUSY}

A dif\/ferent generalization of the creation/annihilation
operators is found in SUSY where, essentially, the term linear in the
coordinate $q$ is replaced by a function of $q$, the so-called
``superpotential'' $W(q)$, leading to the operators
\begin{gather*}
 B^{\pm} = {\frac{1}{\sqrt{2}}} \left[W(q) \mp i  \frac{p_{\rm op}}{\sqrt{m}}\right] .
\end{gather*}
In this case, the term $\omega_0 q$ with constant $\omega_0$ is replaced by a real,
position-dependent function~$W(q)$. The operators $B^{\pm}$ fulf\/il the
commutator and anti-commutator relations
\begin{gather*}
[ B^-,B^+ ]_- = \frac{\hbar}{\sqrt{m}} \frac{dW}{dq}  , \qquad  \{ B^-,B^+ \}_+ =
W^2 + \frac{p_{\rm op}^2}{m}  .
\end{gather*}

The supersymmetric Hamiltonian
\begin{gather*}
H_{\rm SUSY} = \left(\begin{array}{cc} H_1 & 0 \\ 0
      & H_2 \end{array}\right)
\end{gather*}
can be expressed with the help of $B^{\pm}$ in the form
\begin{gather*}
H_1  = B^+ B^-  = -\frac{\hbar^{2}}{2m} \frac{d^{2}}{d q^{2}} + V_1 (q)  ,
\qquad
H_2  = B^- B^+  = -\frac{\hbar^{2}}{2m} \frac{d^{2}}{d q^{2}} + V_2 (q)  .
\end{gather*}

A detailed discussion of the formalism of SUSY can be found, e.g., in~\cite{13}; for our discussion, only the aspects mentioned in the following will
be necessary. Important in this context is that the supersymmetric partner
potentials $V_1(q)$ and $V_2(q)$ fulf\/il real Riccati equations, which follows
directly from the def\/inition of $B^{\pm}$, i.e.,
\begin{gather}
V_1  = \frac{1}{2}  \left[ W^2  - \frac{\hbar}{\sqrt{m}} \frac{dW}{dq} \right]
\label{ric58}
\\
V_2  = \frac{1}{2}  \left[ W^2  + \frac{\hbar}{\sqrt{m}}
  \frac{dW}{dq} \right] .
\label{ric59}
\end{gather}

The energy spectra of $H_1$ and $H_2$ are identical apart from the ground
state. $H_1$ has the ground state $E_0^{(1)} = 0$, whereas the ground state
$E_0^{(2)}$ of $H_2$ is identical with the f\/irst excited state $E_1^{(1)}$ of
$H_1$. The ground state wave function of $H_1$, $\Psi_0^{(1)}$, has no node
and determines the superpotential via
\begin{gather*}
W = - \frac{\hbar}{\sqrt{m}}\frac{\frac{d}{dq}\Psi_0^{(1)}}{\Psi_0^{(1)}}.
\end{gather*}

From equations~\eqref{ric58} and \eqref{ric59}, then, the partner potentials $V_1$ and $V_2$
follow. On the other hand, $\Psi_0^{(1)}$ is connected with $V_1$ via the
solution of the equation $H_1  \Psi_0^{(1)} = 0$, i.e.,
\begin{gather*}
H_1  \Psi_0^{(1)} = -\frac{\hbar^{2}}{2m} \frac{d^{2}}{d
  q^{2}} \Psi_0^{(1)} + \frac{1}{2}  \left[ W^2  - \frac{\hbar}{\sqrt{m}}
  \frac{dW}{dq} \right] \Psi_0^{(1)} =  E_0^{(1)}  \Psi_0^{(1)}  = 0.
\end{gather*}

The connection between the spectra of $H_1$ and $H_2$, i.e.\ $ E_n^{(1)}$ and $
E_n^{(2)}$, and the corresponding wave functions, $\Psi_n^{(1)}$ and $\Psi_n^{(2)}$, is determined
via the generalized creation/annihilation operators~$B^{\pm}$ according to
\begin{gather}
 \Psi_{n+1}^{(1)}  = \frac{1}{\sqrt{E_n^{(2)}}} B^+  \Psi_n^{(2)}
\qquad \mbox{and}\qquad \Psi_n^{(2)} = \frac{1}{\sqrt{E_{n+1}^{(1)}}} B^-  \Psi_{n+1}^{(1)},
\label{ric62}
\end{gather}
where $B^{+}$ creates a node and $B^{-}$ annihilates a node in the wave
function. So, e.g., the f\/irst excited state $\Psi_1^{(1)}$ of $H_1$ (which has one
node) can be obtained from the ground state $\Psi_0^{(2)}$ of $H_2$ (which has no
node) by applying $B^+$ onto it as described in \eqref{ric62}.

In order to obtain the higher eigenvalues and eigenfunctions, $\Psi_0^{(1)}$
in the def\/inition of $W \equiv W_1$ must be replaced by $\Psi_0^{(2)}$,
leading to $W_2  = - \frac{\hbar}{\sqrt{m}} \frac{\frac{d}{dq}\Psi_0^{(2)}}{\Psi_0^{(2)}}$ etc., i.e.,
\begin{gather*}
W_s  = - \frac{\hbar}{\sqrt{m}} \frac{\frac{d}{dq}\Psi_0^{(s)}}{\Psi_0^{(s)}}
\end{gather*}
with the corresponding operators
\begin{gather*}
 B_s^{\pm} = {\frac{1}{\sqrt{2}}}  \left[W_s  \mp   \frac{\hbar}{\sqrt{m}} \frac{d}{dq} \right] ,
\end{gather*}
thus creating a hierarchy that provides all eigenvalues and eigenfunctions of
the Hamiltonians~$H_1$ and~$H_2$.

In the context of this paper, only two systems with exact analytic solutions
shall be considered explicitly, namely, the one-dimensional HO (with constant
frequency $\omega = \omega_0$) and the Coulomb problem. The latter case, a
three-dimensional system with spherical symmetry ($V(\vec{r}) = V(r) = -
\frac{e^2}{r}$), can be reduced to an essentially one-dimensional problem via
separation of radial and angular parts. Using the ansatz $\Phi_{nlm}(\vec{r})
= \frac{1}{r} \Psi_{nl}(r) Y_{lm}(\vartheta,\varphi) = R(r)
Y_{lm}(\vartheta,\varphi)$ for the wave function (with $n$ = total quantum
number, $l$ = azimuthal quantum number, $m$ = magnetic quantum number, $r,
\vartheta, \varphi$ = polar coordinates), the energy eigenvalues $E_n$ of the
system can be obtained from the radial SE
\begin{gather*}
\left\{ -\frac{\hbar^{2}}{2m} \frac{d^{2}}{d r^{2}}  + V_{\rm ef\/f}  \right\}
\Psi_{nl}(r) = E_n  \Psi_{nl}(r)
\end{gather*}
with the ef\/fective potential
\begin{gather}
V_{\rm ef\/f}  = V(r) + \frac{l(l+1)\hbar^2}{2mr^2} = - \frac{e^2}{r} + \frac{l(l+1)\hbar^2}{2mr^2} .
\label{ric66}
\end{gather}

The superpotential $W$, the energy eigenvalues $E_n$ and the supersymmetric
potential $V_1$ for the systems under consideration are given by:

a) HO: $V(q)=\frac{m}{2} \omega^2 q^2$ (eigenfunctions $\Psi_n (q)$: Hermite polynomials)
\begin{gather}
W = \omega q, \nonumber\\ 
E_n  =  \hbar \omega \left( n + \frac{1}{2} \right),\nonumber\\ 
V_1  =  \frac{m}{2} \omega^2 q^2  - \frac{\hbar}{2} \omega  = V(q) - E_0.
\label{ric69}
\end{gather}

b) Coulomb potential: $V(r)= - \frac{e^2}{r}$ (eigenfunctions $\Psi_{nl}(r)$: Laguerre polynomials)
\begin{gather}
W =             \frac{\sqrt{m}
  e^2}{(l+1)\hbar} - \frac{(l+1)\hbar}{\sqrt{m} r}, \nonumber\\ 
E_{n'}  = - \frac{m c^2}{2} \left( \frac{e^2}{\hbar c} \right)^2 \frac{1}{(n'
+ l + 1)^2},\nonumber\\ 
V_1  =  - \frac{e^2}{r} + \frac{l(l+1)\hbar^2}{2mr^2}  + \frac{m c^2}{2}
\left( \frac{e^2}{\hbar c}  \right)^2 \frac{1}{(l + 1)^2} = V_{\rm ef\/f} - E_0 .
\label{ric72}
\end{gather}

In the second case, the radial quantum number $n'$ occurs which indicates
the number of nodes in the wave function and is connected with the total
quantum number $n$, that actually characterizes the energy eigenvalue, via $n =
n' + l + 1$.

Particularly the quantities $V_1$, given in equations~\eqref{ric69} and \eqref{ric72}, shall be
compared with similar expressions obtained in the next subsection where a
nonlinear formulation of time-independent quantum mechanics is presented.

\subsection{Ermakov and nonlinear time-independent quantum mechanics}

In the following discussion of a nonlinear formulation of quantum mechanics,
that is essentially based on the work of Reinisch~\cite{2}, formal
similarities with SUSY (in the time-independent case) and the
complex Riccati formalism (in the time-dependent case) shall be pointed out.

Starting point is Madelung's hydrodynamic formulation of quantum mechanics~\cite{14} that uses the polar ansatz
\begin{gather*}
\Psi (\vec{r},t) = a(\vec{r})  \exp \left\{ - \frac{i}{\hbar} S(\vec{r},t) \right\}
\end{gather*}
for the wave function $\Psi (\vec{r},t)$, turning the linear SE \eqref{ric01} into two
coupled equations for the amplitude $a(\vec{r})$ and the phase $S(\vec{r},t)$,
i.e., the Hamilton--Jacobi-type equation
\begin{gather}
\frac{\partial}{\partial t} S + \frac{1}{2m} (\nabla S)^2
 + V - \frac{\hbar^{2}}{2m} \frac{\Delta a}{a} = 0
\label{ric74}
\end{gather}
(with $\nabla$ = Nabla operator, $\Delta$ = Laplace operator) and the continuity equation
\begin{gather}
\frac{\partial}{\partial t} a^2  +  \frac{1}{m} \nabla (a^2 \nabla S )  = 0,
\label{ric75}
\end{gather}
where $a^2 = \Psi ^* \Psi = \varrho (\vec{r},t)$ is the usual probability
density. For stationary states, the energy of the system is related with the
action $S$ via $\frac{\partial}{\partial t} S = -E ={\rm const}$ and the density is time-independent, i.e.,
\begin{gather*}
\frac{\partial}{\partial t} a^2 =0,
\end{gather*}
where it subsequently follows that the second term on the lhs of equation~\eqref{ric75} must
also vanish. In the usual textbook treatment, this is achieved by taking $\nabla S
= 0$, thus
turning equation~\eqref{ric74} into the conventional time-independent linear SE
\begin{gather}
- \frac{\hbar^{2}}{2m} \Delta a  + V a  = E a.
\label{ric77}
\end{gather}

This is, however, not the only possibility of fulf\/illing equation~\eqref{ric75} since, also for $\nabla S
\neq 0$, this can be accomplished if only
\begin{gather}
\nabla S  = \frac{C}{a^2}
\label{ric78}
\end{gather}
with constant $C$. In this case, equation~\eqref{ric74} takes the form of a nonlinear Ermakov equation
\begin{gather}
\Delta a +\frac{2m}{\hbar} (E - V) a  =  \left(\frac{C}{\hbar} \right)^2
\frac{1}{a^3} .
\label{ric79}
\end{gather}

The corresponding complex Riccati equation, equivalent to equation~\eqref{ric04} in the
time-depen\-dent case, is given here by
\begin{gather}
\nabla \left( \frac{\nabla \Psi}{\Psi} \right)  + \left( \frac{\nabla
    \Psi}{\Psi} \right)^2  +  \frac{2m}{\hbar} (E - V)  = 0,
\label{ric80}
\end{gather}
where the following substitutions must be made
\begin{gather*}
\frac{\partial}{\partial t}  \leftrightarrow  \nabla
, \qquad \left(\frac{2\hbar}{m}y\right) = \frac{\dot{\lambda}}{\lambda}
\leftrightarrow   \frac{\nabla \Psi}{\Psi}  ,\qquad \lambda = \alpha e^{i\varphi}
 \leftrightarrow  \Psi = a e^{i \frac{S}{\hbar}}  .
\end{gather*}

Considering f\/irst the one-dimensional HO, and introducing the dimensionless
variable $\zeta$ via $\zeta = |k_0| q$ with $\hbar k_0 = p_0 = \pm
\sqrt{2mE}$, $\tilde{V}(\zeta) = V[q(\zeta)]$ and $\ddot{a} = \frac{d^2}{d
  \zeta^2} a$, equation~\eqref{ric79} acquires the familiar form
\begin{gather}
\ddot{a} + \left( 1 - \frac{\tilde{V}}{E} \right) a  = \frac{1}{a^3}  .
\label{ric82}
\end{gather}

A similar formulation of the time-independent SE in terms of this equation,
but within a~dif\/ferent context and dif\/ferent applications has also been given
in \cite{15}. In another paper~\cite{16} the relation between the Ermakov
equation~\eqref{ric82} and the time-independent SE has been extended to also include
magnetic f\/ield ef\/fects. The nonlinear dif\/ferential equation~\eqref{ric82} has also been
used to obtain numerical solutions of the time-independent SE for single and
double-minimum potentials as well as for complex energy resonance states; for
details see~\cite{17}. Here we want to concentrate on the similarities between
the time-dependent and time-independent situation, in particular with respect
to SUSY, as mentioned in Sections~5.1 and~5.2.

Following the method described in~\cite{2}, from the {\it real} solution
$a_{\rm NL}(\zeta)$ of this {\it nonlinear} Ermakov equation~\eqref{ric82} the {\it complex}
solution $a_{\rm L} (\zeta)$ of the {\it linear} SE \eqref{ric77} can be obtained via
\begin{gather}
a_{\rm L} (\zeta) = a_{\rm NL}(\zeta)  \exp \left\{ - \frac{i}{\hbar} S
\right\} = a_{\rm NL}(\zeta)  \exp \left\{ - i \int^{\zeta}_{\zeta_0} d\zeta '
  \frac{1}{a_{\rm NL}^2 (\zeta ')} \right\}  ,
\label{ric83}
\end{gather}
from which a {\it real} (not normalized) solution of the time-independent SE can be
constructed according to
\begin{gather}
\tilde{a}_{\rm L} (\zeta) =  \Re [a_{\rm L} (\zeta)]  = \frac{1}{2} \left[ a_{\rm NL}
 e^{\frac{i}{\hbar} S} +  a_{\rm NL}  e^{- \frac{i}{\hbar} S}
\right] = a_{\rm NL} \cos \left( \int^{\zeta}_{\zeta_0} d\zeta '\frac{1}{a_{\rm NL}^2 (\zeta ')} \right)  .
\label{ric84}
\end{gather}

So far, the energy $E$ occurring in equation~\eqref{ric82} is still a free parameter that
can take any value. However, solving equation~\eqref{ric82} numerically for arbitrary values
of $E$ leads, in general, to solutions $a_{\rm NL}$ that diverge for increasing
$\zeta$. Only if the energy $E$ is appropriately tuned to any eigenvalue $E_n$ of
equation~\eqref{ric77} does this divergence disappear and the integral in the cosine of
equation~\eqref{ric84} takes for $\zeta \rightarrow \infty$ exactly the value $\frac{\pi}{2}$, i.e., the cosine vanishes at
inf\/inity. So, the quantization condition that is usually obtained from the
requirement of the truncation of an inf\/inite series in order to avoid
divergence of the wave function is, in this case, obtained from the
requirement of nondiverging solutions of the nonlinear Ermakov equation~\eqref{ric82}
by variation of the parameter~$E$. This has been numerically verif\/ied in the
case of the one-dimensional HO and the Coulomb problem and there is the conjecture that
this property is ``universal'' in the sense that it does not depend on the
potential $V$ (see~\cite{2}).

For comparison with the situation in SUSY, the HO and the Coulomb problem can
be written in the form:

a) HO: with $\mu = \left( \frac{\hbar \omega}{2 E} \right)^2$, $E = E_n = (n +
\frac{1}{2} ) \hbar \omega \rightarrow \mu_n = \frac{1}{(2n + 1)^2}$ and $\mu
\zeta^2 = \frac{\frac{m}{2} \omega^2 q^2}{E} = \frac{V}{E}$ follows:
\begin{gather}
\ddot{a} + \left( 1 - \mu \zeta^2  \right) a  = \ddot{a} + \left( 1 -
  \frac{\tilde{V}}{E} \right) a = \ddot{a} -  \frac{U_n}{E} a  = \frac{1}{a^3}  ,
\label{ric85}
\end{gather}
where
\begin{gather*}
U_n  =  \frac{m}{2} \omega^2 q^2  - \hbar \omega \left( n + \frac{1}{2}
\right)  = V(q) - E_n .
\end{gather*}

b) Coulomb problem: with $a(r,\vartheta,\varphi) = R(r)
Y_{lm}(\vartheta,\varphi)$ the radial part can be separated and,
with the dimensionless variable $\zeta = | k_0 | r$ with now $\hbar k_0 = p_0 = \pm
\sqrt{2m(-E)}$   $(E < 0)$, the radial wave function
can be written as $X(\zeta) = r(\zeta) X[r(\zeta)]$, which corresponds to $\Psi_{nl}(r)$ in SUSY. This function
obeys, again, an Ermakov equation, namely
\begin{gather}
\ddot{X} + \left( \frac{\tilde{W}}{E} - 1 \right) X
 = \ddot{X} + \frac{U_ {n'}}{E} X = \frac{1}{X^3}  ,
\label{ric87}
\end{gather}
where
\begin{gather*}
\tilde{W}(\zeta)  = \tilde{V}[r(\zeta)] + \frac{l(l+1)
  \hbar^2}{2mr^2(\zeta)} \; \hat{=} \; V_{\rm ef\/f}
\end{gather*}
is just the ef\/fective potential $V_{\rm ef\/f}$ from SUSY (see equations~\eqref{ric66} and \eqref{ric72}) and
$E = E_n = - \frac{m e^4}{2 \hbar^2 n^2}$ with $n =
n' + l + 1 $. The coef\/f\/icient of the term linear in $X$ can again be
expressed with the help of the potential-like expression $U_{n'}$ as
\begin{gather*}
U_{n'}  =  - \frac{e^2}{r} + \frac{l(l+1)\hbar^2}{2mr^2}  + \frac{m c^2}{2}
\left( \frac{e^2}{\hbar c}  \right)^2 \frac{1}{(n' + l + 1)^2} = V_{\rm ef\/f} - E_{n'} .
\end{gather*}

In both cases, the {\it ground state} $(n = 0)$ wave functions are real,
nodeless $(n' = 0)$ and the phase does not depend on spatial variables (i.e.,
$\nabla S
= 0$). Therefore,
the rhs of equations~\eqref{ric85} and \eqref{ric87} vanishes since $\frac{1}{a^3} \propto (\nabla
S)^2 a$ (similar for the Coulomb
problem), i.e., the nonlinear Ermakov equations turn into the usual
time-independent SEs. In this case, comparison shows that for the HO and the
Coulomb problem, the potential-like terms $U_0$ are identical with the
corresponding $V_1$ of SUSY. For $n > 0$ and $n' > 0$, however, $U_n$ and $U_{n'}$ are dif\/ferent
from $V_1$ and describe higher excited states. In SUSY, these states can only
be obtained from the hierarchy described in the previous subsection. Here, the
price that must be paid to include these excited states is the nonlinearity on
the rhs of equations~\eqref{ric85} and~\eqref{ric87}.

Comparing the situation in this nonlinear formulation of time-independent
quantum mechanics with SUSY and the time-dependent systems discussed in the
f\/irst part of this paper, one can see the following similarities:

i) {\it Comparison with SUSY}:

The {\it real} superpotential $W = - \frac{\hbar}{\sqrt{m}} \left( \frac{\nabla
    \Psi_0}{\Psi_0} \right)$ is replaced by the {\it complex}
``superpotential'' $\mathcal{C} (q) = - \frac{\hbar}{\sqrt{m}} \left[ \left( \frac{\nabla
    |\Psi|}{|\Psi|} \right) + i \frac{\nabla S}{\hbar}
\right]$, i.e., the ground state $\Psi_0$ is replaced by the
absolute value $|\Psi|$ of any eigenstate and an additional imaginary part
depending on the phase $\frac{S}{\hbar}$ of the wave function occurs, being responsible for
the non-vanishing rhs of the Ermakov equations~\eqref{ric85} and~\eqref{ric87}.

ii) {\it Comparison with time-dependent SE}:

As mentioned before, the complex quantity $\mathcal{C} (q)$ can be compared with the
time-dependent quantity $\left( \frac{2\hbar}{m} y \right)$ fulf\/illing the complex Riccati equation (4),
$\mathcal{C} (q) = - \frac{\hbar}{\sqrt{m}} \frac{\nabla
    \Psi}{\Psi} \leftrightarrow \left( \frac{2\hbar}{m} y(t)  \right) =
  \frac{\dot\lambda}{\lambda}$, or, in terms of real and imaginary parts, $\frac{\nabla
    |\Psi|}{|\Psi|} \leftrightarrow \frac{\dot{\alpha}}{\alpha}$ and $\frac{\nabla S}{\hbar} \propto \frac{1}{
    |\Psi|^2}  \leftrightarrow \dot{\varphi} \propto \frac{1}{\alpha^2}$.

\section{Conclusions and perspectives}

As a result of this investigation, one can state that the Ermakov invariant is
a quantity of central importance that connects dif\/ferent forms for the
description of the dynamics of quantum systems, such as the time-dependent SE,
the time-dependent Green's function (or Feynman kernel) and the
time-dependent Wigner function. Unlike the classical Hamiltonian or
Lagrangian, this invariant not only depends on the classical variables such as
position and momentum, but also on the quantum uncertainties of these
quantities contained in $\alpha$ and $\dot{\alpha}$. Therefore, the initial values of these
quantities also play an important role for the time-evolution of the quantum
system, as has been demonstrated in the discussion of the time-dependence of
the WP width or position uncertainty. So, the time-evolution of a typical
quantum mechanical property can be totally described if one only knows the
{\it classical} trajectory $\eta$ and the {\it classical} velocity $\dot{\eta}$ (including their
initial conditions) plus the {\it initial} position {\it uncertainty}. This traces
quantum dynamics entirely back to the classical one plus the existence of an
uncertainty principle.

So far, the discussion of time-dependent systems included only systems where the potential is at most
quadratic in its variables (a similar treatment of the motion in a magnetic
f\/ield is also possible; see, e.g.,~\cite{18}). This might not be as
restrictive as it seems at f\/irst sight since one may sometimes perform
canonical transformations to reduce a given Hamiltonian to a quadratic form
\cite{19} which has been shown explicitly by Sarlet for some polynomial
Hamiltonians.  To what extent this method can also be applied in our case requires
more detailed studies. Another way to extend considerably the class of Hamiltonians for
which an exact invariant can be found is by making use of generalized
canonical transformations~\cite{20}. Further generalizations of applying the
Ermakov-type invariants also in the context of f\/ield-atom interactions,
systems of coupled oscillators including damping and/or time-dependent masses
and attempts of obtaining the corresponding Wigner functions can be found in~\cite{21}. For a further survey of two-dimensional problems in this context,
see also~\cite{22}.

Similar to the factorization of the Hamiltonian of the HO in
terms of (classical) complex normal modes, or, (quantum mechanical) in terms
of creation and annihilation operators $b^{\pm}$, a factorization of the Ermakov invariant is possible that looks like a
complex time-dependent generalization of this formalism. The major dif\/ference
compared with the conventional case is the replacement of $\pm i \omega_0$ in
front of the spatial variable $q$ in equation~\eqref{ric51} by the complex time-dependent
quantity $\left( \frac{2\hbar}{m} y  \right)$, which fulf\/ils the Riccati equation \eqref{ric04}, or its conjugate
complex~$\left( \frac{2\hbar}{m} y^*  \right)$, respectively.

Another generalization of the creation/annihilation operator formalism of the
HO, concerning the space-dependence instead of the time-dependence, is found
in SUSY where the term linear in the spatial coordinate $q$ is replaced by a
function of $q$, the so-called ``superpotential''~$W(q)$. The generalized
creation and annihilation operators $B^{\pm}$ are obtained by replacing the
term $\omega_0 q$ in $b^{\pm}$ by (up to constants) $W(q)$, where $W(q)$
fulf\/ils the {\it real} Riccati equations~\eqref{ric58} and~\eqref{ric59}.

In a nonlinear formulation of time-independent quantum mechanics proposed by
Reinisch~\cite{2}, the amplitude of the wave function fulf\/ils a real
nonlinear (space-dependent) Ermakov equation, that can be connected with the
time-independent SE written as a {\it complex} space-dependent Riccati equation
(see~\eqref{ric80}) together with a kind of conservation law (see \eqref{ric78}), similar to the
conservation of ``angular momentum'' in the complex plane (see \eqref{ric15}) in the
time-dependent case. In SUSY, the superpotential is initially determined by
the amplitude of the ground state wave function (i.e., without any
contribution from the phase) and the excited states can be obtained by some
hierarchy based thereupon, in the nonlinear formulation of time-independent
quantum mechanics,the ground state wave function is replaced by the absolute
value of any eigenstate plus an imaginary part depending on the gradient of
the phase of this state. Similar to the time-dependent situation, this looks
like a complex generalization where, now, the real superpotential~$W(q)$ is
replaced by a complex term that not only depends on the amplitude of the wave
function, but, due to an additional imaginary contribution, also on (the
gradient of) its phase. Further clarif\/ication of these facts will be subject
of forthcoming studies.

\subsection*{Acknowledgements}

The author wishes to thank G.~Reinisch for numerous encouraging and inspiring discussions.

\pdfbookmark[1]{References}{ref}
\LastPageEnding

\end{document}